\documentclass[11pt,twoside]{article}

\usepackage{asp2014}

\aspSuppressVolSlug
\resetcounters

\newcommand{\helix}[1]{{H\tiny{E}\normalsize{LI}\tiny{X}$^{{\textstyle+}}$#1}} 
\newcommand{\bm}[1]{\mbox{\boldmath$#1$\unboldmath}}

\begin{document}

\title{Performance analysis of the SO/PHI software framework for on-board data reduction}

\author{K. Albert,$^1$ J. Hirzberger,$^1$ D. Busse,$^1$ J. Blanco Rodr\'{i}guez,$^2$, J. S. Castellanos Dur\'{a}n$^1$, J. P. Cobos Carrascosa,$^3$ B. Fiethe,$^4$ A. Gandorfer,$^1$ Y. Guan,$^4$ M. Kolleck,$^1$ A. Lagg,$^1$ T. Lange,$^4$ H. Michalik,$^4$ S. K. Solanki,$^1$  J. C. del Toro Iniesta,$^3$ and J. Woch$^1$}
\affil{$^1$Max Planck Institute for Solar System Research, G\"{o}ttingen, Germany,\\e-mail: albert@mps.mpg.de,\\ $^2$Universidad de Valencia, Paterna (Valencia), Spain\\ $^3$Instituto de Astrof\'{i}sica de Andaluc\'{i}a (IAA - CSIC), Granada, Spain\\ $^4$Institute of Computer and Network Engineering, TU Braunschweig, Germany}

\begin{abstract}
The Polarimetric and Helioseismic Imager (PHI) is the first deep-space solar spectropolarimeter, on-board the Solar Orbiter (SO) space mission. It faces: stringent requirements on science data accuracy, a dynamic environment, and severe limitations on telemetry volume. SO/PHI overcomes these restrictions through on-board instrument calibration and science data reduction, using dedicated firmware in FPGAs.\\
This contribution analyses the accuracy of a data processing pipeline by comparing the results obtained with SO/PHI hardware to a reference from a ground computer. The results show that for the analysed pipeline the error introduced by the firmware implementation is well below the requirements of SO/PHI.
\end{abstract}

\section{Introduction}
The Polarimetric and Helioseismic Imager (PHI) is one of ten instruments to orbit the Sun on-board Solar Orbiter \citep[SO; see][]{Mueller2013_SO}. SO/PHI \citep{Solanki_PHI}, is an imaging spectropolarimeter, probing the photospheric Fe\,I\,6173\,$\mbox{\AA}$ absorption line. 

SO/PHI records data in five dimensions: time series of data sets containing $2048 \times 2048$\,pixel images of the Sun, sampling the target absorption line at six wavelengths, recording four different polarisation states at each wavelength. These polarisation states contain linear combinations of the Stokes parameters ($\mathbf{S}=[I,Q,U,V]^T$), a formalism to describe the polarisation of light in terms of four ideal polarisation filters. To arrive to the Stokes images (the input for scientific analysis), the recorded polarisation states are demodulated with the demodulation matrix. These images, complemented with a wavelength dimension, encode the magnetic field vector at the mean formation height of the absorption line and the line of sight (LOS) velocity due to the Zeeman and Doppler effects. Arriving to these quantities is possible by the inversion of the Radiative Transfer Equation (RTE). See \citet{Iniesta2003introduction} for more details on spectropolarimetry.

SO/PHI is the first spectropolarimeter on a deep space mission, facing an unprecedented dynamic environment and telemetry limitations. These challenges are met with a full and autonomous on-board data analysis system: it determines the instrument characteristics, applies them to the science data, then derives the targeted physical parameters. This system is implemented on a data processing unit with two Field Programmable Gate Arrays (FPGAs), reconfigured in flight to perform image processing functions \citep{fiethe2012adaptive, lange2017board}, and a microprocessor running a data processing framework that combines these functions into pipelines \citep{Albert2018_Autonomous}. This contribution analyses errors induced by the on-board processing.

\section{The on-board data analysis software}
The science data processing comprises of preprocessing and RTE inversion. The preprocessing primarily corrects the images for the dark and flat field of the instrument, and does the polarimetric demodulation. Depending on science case and the instrument parameters determined at instrument commissioning, it may have additional steps (e.g. spatial cropping or deconvolution from image artefacts). The RTE inversion transforms the 24-image spectropolarimetric dataset into 5 images of interest: azimuth, inclination and magnitude of the magnetic field, the LOS velocity and the total intensity at continuum wavelength. See \citet{carrascosa2016rte} for details on SO/PHI's RTE inversion scheme. To save FPGA resources, the preprocessing functions use fixed point number representation on 24.8 bits, while the RTE inversion is on 32 bits floating point.

The most basic preprocessing pipeline for a data set from an imaging spectropolarimeter contains dark and flat field correction and polarimetric demodulation:
\begin{equation}
 \bm{S}_{\lambda}(x,y) = D(x,y) \cdot [(\bm{I}^{obs}_{\lambda}(x,y) - I^{dark}(x,y))/I^{flat}(x,y)],
\label{eq:pipeline}\end{equation}where "$\cdot$" denotes matrix multiplication, $\lambda$ marks wavelength dependence, $x$ and $y$ are spatial dimensions. The Stokes parameters are noted with \(\bm{S}\), $D$ is the $4 \times 4$ Demodulation Matrix, $\bm{I}^{obs}$ is the observed data in the four modulation states. The dark field of the sensor is $I^{dark}$, while $I^{flat}$ is the flat field of the telescope, neither depending on wavelengths and modulation states (may change for the flat field after commissioning).

\begin{figure}[t]
\centering
\includegraphics[trim={5mm 5mm 5mm 5mm},clip,width=0.47\textwidth]{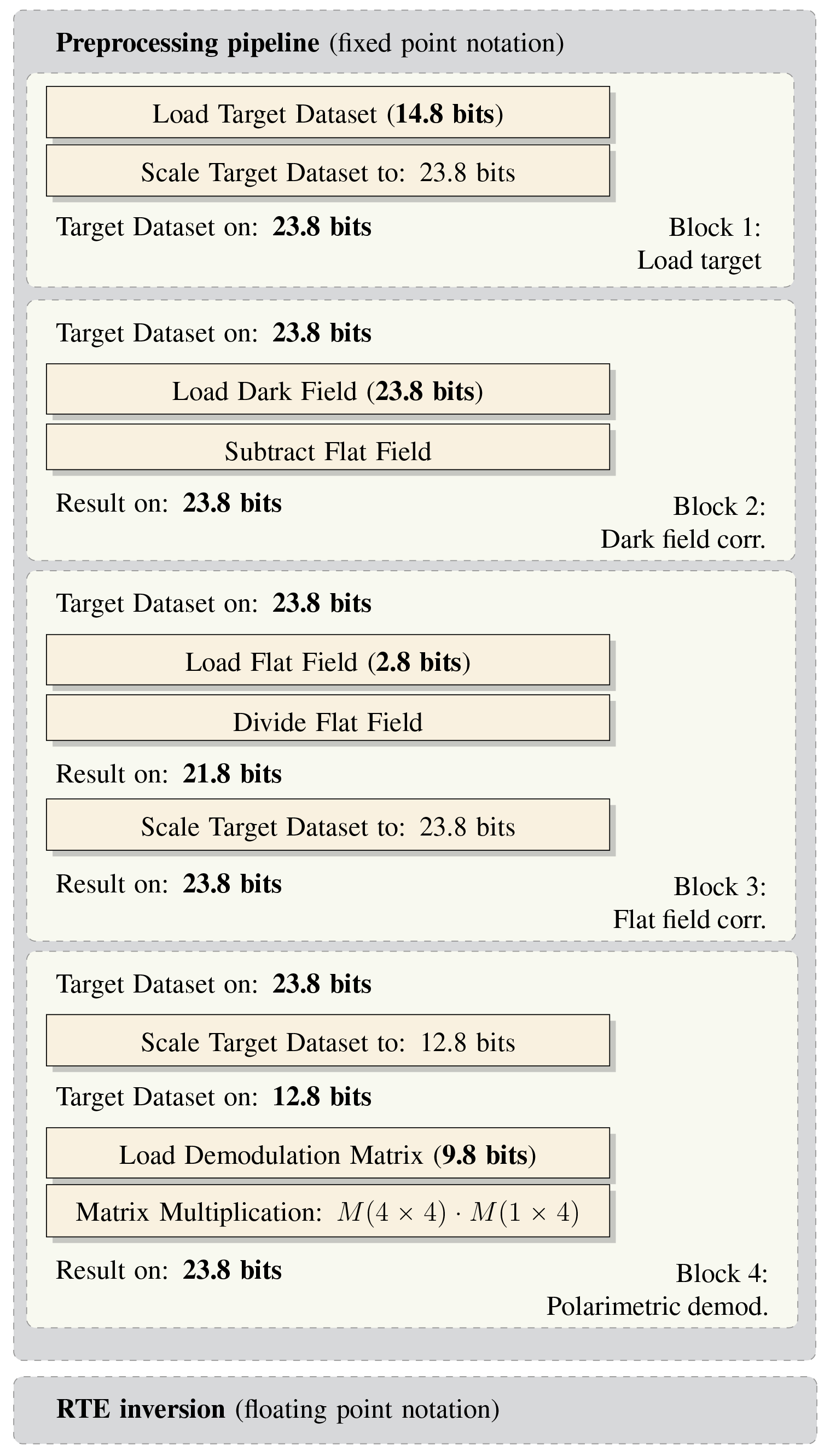}
\caption{The studied pipeline. The preprocessing controls accuracy by scaling.}
\label{fig:Pipeline} 
\end{figure}

To implement eq. \ref{eq:pipeline}, we use four blocks, combined into a pipeline (see fig. \ref{fig:Pipeline}). The raw data is integer, represented up to 22.8 bits after accumulation (14.8 assumed in test). As the exposure time is calibrated to fill a defined percentage of the detector full well, the recorded data is ideally represented. To process the data at the highest resolution, we shift the pixel values of these images to the top of the full range ($\cdot 2^9$), arriving to 23.8 bits (one bit is sign). This representation is the block interface, however some blocks re-scale the images to optimise the output accuracy.

We quantify the errors introduced through on-board processing by running the pipeline on a SO/PHI ground reference model and on ground computer (using floating point in Python). The test data is from the Solar Dynamics Observatory / Helioseismic and Magnetic Imager \citep{Schou2012HMI}, run through the SO/PHI instrument simulator, SOPHISM \citep{Blanco_SOPHISM}. This data is modulated into measured intensities, then degraded with the ground-measured flat and dark field of the SO/PHI flight model (also used by the pipeline). We compare the results of the preprocessing after flat field correction, and polarimetric demodulation (no errors are expected from subtraction). The RTE inversion is done on a ground computer, with the He-Line Information Extractor inversion code \citep[\helix{}; see][]{Lagg2004}. \helix{} assumes a Milne-Eddington approximation of the atmosphere, the same as SO/PHI's on-board inversion scheme, however the profile fitting method is different. The differences in physical parameters only indicate the magnitude of the error expected from the numerical inaccuracies in the preprocessing, due to the differences in the algorithms, the innate uncertainty in the results of the inversion, but also the nature of profile fitting.

\section{Results}
The errors from the division are below $10^{-3}$ (compared to the reference results), apart from a few outliers in the divisor, with Root Mean Square (RMS) around $5\cdot10^{-5}$.

\begin{figure}[t]
	\centering
	\includegraphics[width=0.5\textwidth]{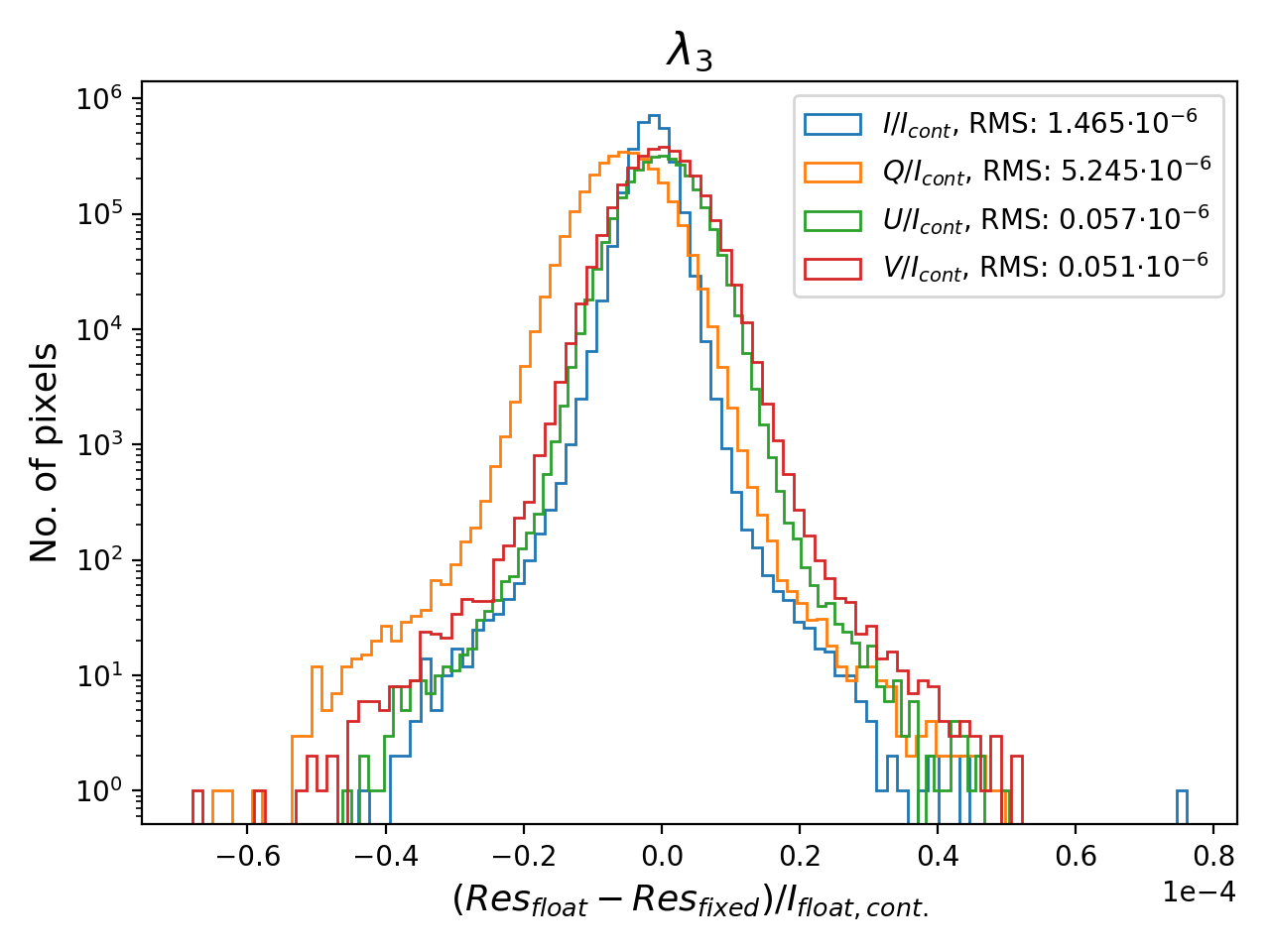}
	\caption{Histogram of polarimetric errors, showing requirement compliance.}
	\label{fig:Hist_Err_Dem}
\end{figure}

SO/PHI requires the accuracy of the polarisation signals (i.e. $\bm{S}$) to be better than $10^{-3}$. Fig. \ref{fig:Hist_Err_Dem} shows the error histogram at one wavelength sample. All pixels comply, with their RMS in the order of $10^{-6}$, leaving a large margin for other error sources. The errors decrease from the previous step due to the nature of polarimetry: it calculates the difference between signals, partially cancelling previous errors. Furthermore, the error in $Q$ is larger than in the rest of the Stokes images, due to a small term in $D$.

The RTE inversion errors show only an indication of expected error from numerical inaccuracies. The error RMS in magnetic field strength error, azimuth, inclination (calculated in the sunspot) and LOS velocity (calculated in the solar disk) introduced by the calculation errors are 33.64\,G, 1.92$^\circ$, 2.56$^\circ$, and 19\,ms$^{-1}$, respectively. The inversion process for this type of data set, calculated statistically, has error RMS 21.9\,G, 1.37$^\circ$, 1.34$^\circ$ and 14.5\,ms$^{-1}$, respectively. What is introduced on top of this by the numerical inaccuracies amount to 53\%, 40\%, 91\% and 31\% of the inversion error.

\section{Conclusions}
SO/PHI is the first instrument of its kind to perform on-board data analysis, including data preprocessing and the inversion of the RTE. These steps use computationally demanding image processing functions, implemented on FPGAs. The fixed point number representation in the on-board preprocessing was motivated by resource limitations.

The errors induced by the preprocessing conform with requirements, with a good margin for other sources. This is achieved by keeping full control over data accuracy, a significant overhead. Errors in Fourier domain processing are currently being analysed.

\section*{Acknowledgements}
Workframe: International Max Planck Research School (IMPRS) for Solar System Science. Solar Orbiter: ESA, NASA. Support grants: DLR 50 OT 1201, Spanish Research Agency ESP2016-77548-C5, European FEDER. Data: NASA/SDO HMI science team.

\bibliography{P4-1}  

\end{document}